\documentclass{emulateapj}
\usepackage{textcomp}
\usepackage{mathtext}
\usepackage{amssymb}

\newcommand{\fermi}{\textit{Fermi}}
\newcommand{\gr}{$\gamma$-ray}
\newcommand{\psr}{PSR J1311$-$3430}

\begin{document}

\title{Discovery of Gamma-Ray Orbital Modulation in the black widow PSR J1311$-$3430}

\author{Yi Xing and Zhongxiang Wang}

\affil{Key Laboratory for Research in Galaxies and Cosmology,
Shanghai Astronomical Observatory,\\ 
Chinese Academy of Sciences, 80 Nandan Road, Shanghai 200030, China}

\begin{abstract}
We report our discovery of orbitally modulated \gr\ emission from the black
widow system \psr. We analyze the \textit{Fermi} Large Area Telescope data
during the offpulse phase interval of the pulsar,
and find the orbital modulation signal at a $\sim$3$\sigma$ confidence level.
Further spectral analysis shows no significant 
differences for the spectra obtained during the bright and faint orbital 
phase ranges. A simple sinusoid-like function can describe the modulation.
Given these properties, we suggest that the intrabinary \gr\ emission
arises from the region close to the companion and the modulation is caused
by the occultation of the emitting region by the companion, similar to that
is seen in the transitional millisecond pulsar binary (MSP) PSR J1023+0038.
Considering the X-ray detection of intrabinary shock emission from eclipsing 
MSP binaries recently reported, this discovery further suggests 
the general existence of intrabinary \gr\  emission from them.

\end{abstract}


\section{Introduction}

Millisecond pulsars (MSPs) are widely accepted to be old neutron stars that
were spun up through mass accretion from the companions when 
they were at the low-mass X-ray binary phase \citep{alp+82,rs82}. 
Not surprisingly,
$>$60\% known MSPs are in binaries (e.g., \citealt{man+05}).
A sub-class of them, so-called `black widow' pulsar systems \citep{fst88},
have very low-mass, $\sim$0.02\,$M_{\sun}$ companions.
To form isolated MSPs, one possible channel is through ablation of
the companions by the pulsar wind. This possibility likely occurs in  
the black widows since they are eclipsing systems at radio frequencies,
indicating the interaction between the pulsar wind and the companions.
X-ray observations of them revealed orbital flux variations, also supporting
the presence of the intrabinary interaction \citep{hua+12,gen+14}.
In addition, recent extensive studies of the so-called `redback' systems
\citep{r13}
have provided clear evidence for the interaction. These redbacks are also
eclipsing MSP binaries, but contain relative massive, 
$\sim$0.1--0.6\,$M_{\sun}$ 
companions. X-ray observations of the prototypical redback PSR J1023+0038
detected significant orbital flux variations \citep{arc+10,bog+11}, and 
the variations can be explained by the existence of an intrabinary shock
region \citep{bog+11}. Similar features were also clearly seen in
the redback XSS J12270$-$4859 (\citealt{bog+14} and references therein). 

Owing to its all-sky monitoring and high sensitivity capabilities,
the \fermi\ Gamma-ray
Space Telescope, launched in 2008, has greatly improved our studies of 
pulsars. For MSPs,
more than six-fold black widows and redbacks have been discovered with
the help of \fermi\ (e.g., \citealt{r13}). At \fermi's 100\,MeV to 300\,GeV 
energy range, marginal evidence for the intrabinary interaction in
the eclipsing systems has also been seen. For the first discovered black widow
PSR B1957$+$20 \citep{fst88}, an orbital modulation signal was detected
at a $\sim 2.3\sigma$ confidence level \citep{wu+12}. In addition, possible
signals were also reported for XSS J12270$-$4859 \citep{xw14} and
a candidate redback 2FGL J0523.3$-$2530 \citep{xwn14}.
Theoretical studies have long predicted the intrabinary interaction and
related high-energy emission from black widows (e.g., \citealt{at93}).
Studies of the \gr\ emission from the intrabinary region allow us to explore 
the detailed physical processes within such a binary (e.g., \citealt{rob+14}).
In this paper we report the detection of orbitally modulated \gr\ emission 
from a recently discovered black widow \psr, which thus indicates
the intrabinary origin for part of its emission.

\psr\ was initially listed as an unassociated source in the \fermi\ Large
Area Telescope (LAT) source catalog (2FGL J1311.7$-$3429; \citealt{nol+12}). 
It is the only \gr\ selected MSP with \gr\ pulsed emission discovered 
via a direct blind search in the \fermi\ data \citep{ple+12}.
The pulsed radio emission soon was detected too \citep{ray+13}, but
the signal was visible only during $<$10\% of the observation time, suggesting
strong variations in the intrabinary medium.
Before the discovery, the source was found to have orbital 
modulation with a short period of $\simeq$94 minutes through optical imaging 
and spectroscopy \citep{r12,rom+12}. Considering its properties of
weak X-ray emission, sinusoid-like optical modulation and large modulation
amplitude, it was already suggested to be a black widow system 
and the optical modulation is caused by irradiation of the companion
by the pulsar wind \citep{r12}. 
Marginal modulated X-ray emission possibly related to the intrabinary shock 
has also been detected \citep{r12,kat+12}. 
The \gr\ discovery of the 2.56 ms spin signal thus confirmed its black 
widow nature \citep{ple+12}. 
Analyzing the \fermi\ data, we searched and found the orbital modulation 
signal from the source's
offpulse emission. Below we present the data analysis and results in
Section~\ref{sec:ana}. The results are discussed in Section~\ref{sec:dis}.
\begin{center}
\includegraphics[scale=0.2]{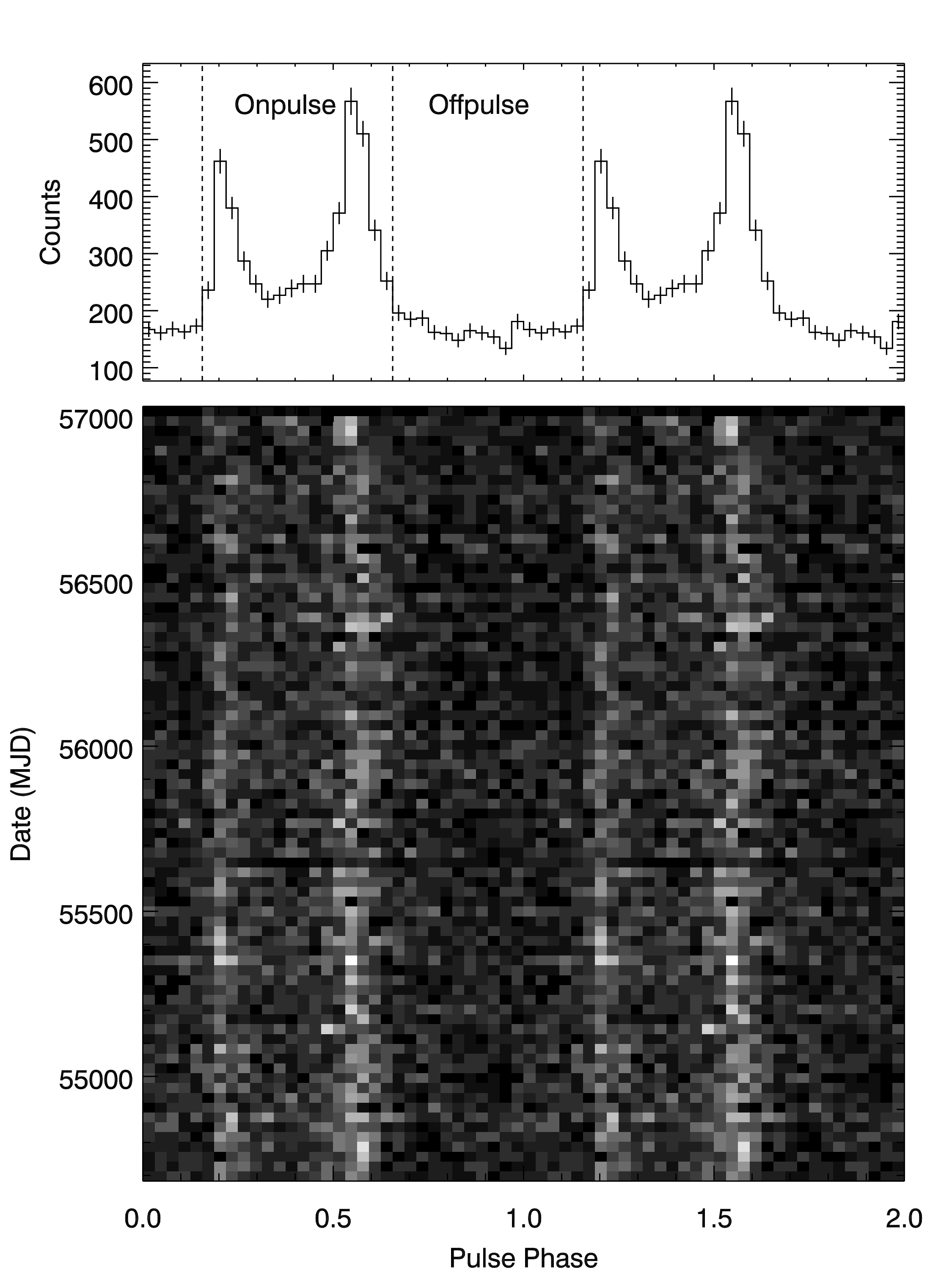}
\figcaption{Folded pulse profile and two-dimensional phaseogram in 
32 phase bins obtained for \psr. The gray scale represents the number 
of photons in each bin, and the dashed lines mark the onpulse and
offpulse phase intervals.
\label{fig:timing}}
\end{center}

\section{Data Analysis and Results}   
\label{sec:ana}

\subsection{\fermi\ LAT data}
\label{subsec:data}

We selected 0.1--300 GeV LAT events from the \textit{Fermi} Pass 
7 Reprocessed (P7REP) database inside a $\mathrm{20^{o}\times20^{o}}$ 
region centered at the position of \psr\ \citep{ple+12} during the time 
period from 2008-08-04 15:43:36 to 2015-01-06 21:19:57 (UTC). 
Only events with zenith angle less than 100\,deg  
and during good time intervals were kept. The former prevents 
the Earth's limb contamination, and for the latter, the quality of 
the data was not affected by the spacecraft events. 

\subsection{Timing Analysis}
\label{subsec:ta}

We performed timing analysis to the 0.1--300 GeV LAT data of 
the \psr\ region to update the \gr\ ephemeris given in \citet{ple+12}. 
An aperture radius of 1\fdg0 was used. We determined the pulse time of 
arrivals (TOAs) by obtaining the pulse profiles of 40 evenly divided 
segments using the known ephemeris \citep{ple+12} and cross-correlated 
them with a template profile created with data during the time period 
of MJD 54682--56119 (the same time range as that in \citealt{ple+12}), 
following the algorithm described in \citet{t92}. We used 
TEMPO2 \citep{hem06,ehm06} to fit the TOAs. Only pulse frequency $f$ 
and frequency derivative $\dot{f}$ were fitted, and the other timing 
parameters were fixed to their known values. We obtained 
$f= 390.56839326403(7)$\,Hz and $\dot{f}= -3.193(1)\times 10^{-15}$\,s$^{-2}$,
consistent with the values given in \citet{ple+12} within $\sim$0.5$\sigma$ 
and $\sim$2.2$\sigma$ uncertainties, respectively. The folded pulse profile 
and two-dimensional phaseogram are shown in Figure~\ref{fig:timing}. 
We defined phase 0.16--0.66 and 0.66--1.16 
as the onpulse and offpulse phase intervals, respectively.
\begin{center}
\includegraphics[scale=0.32]{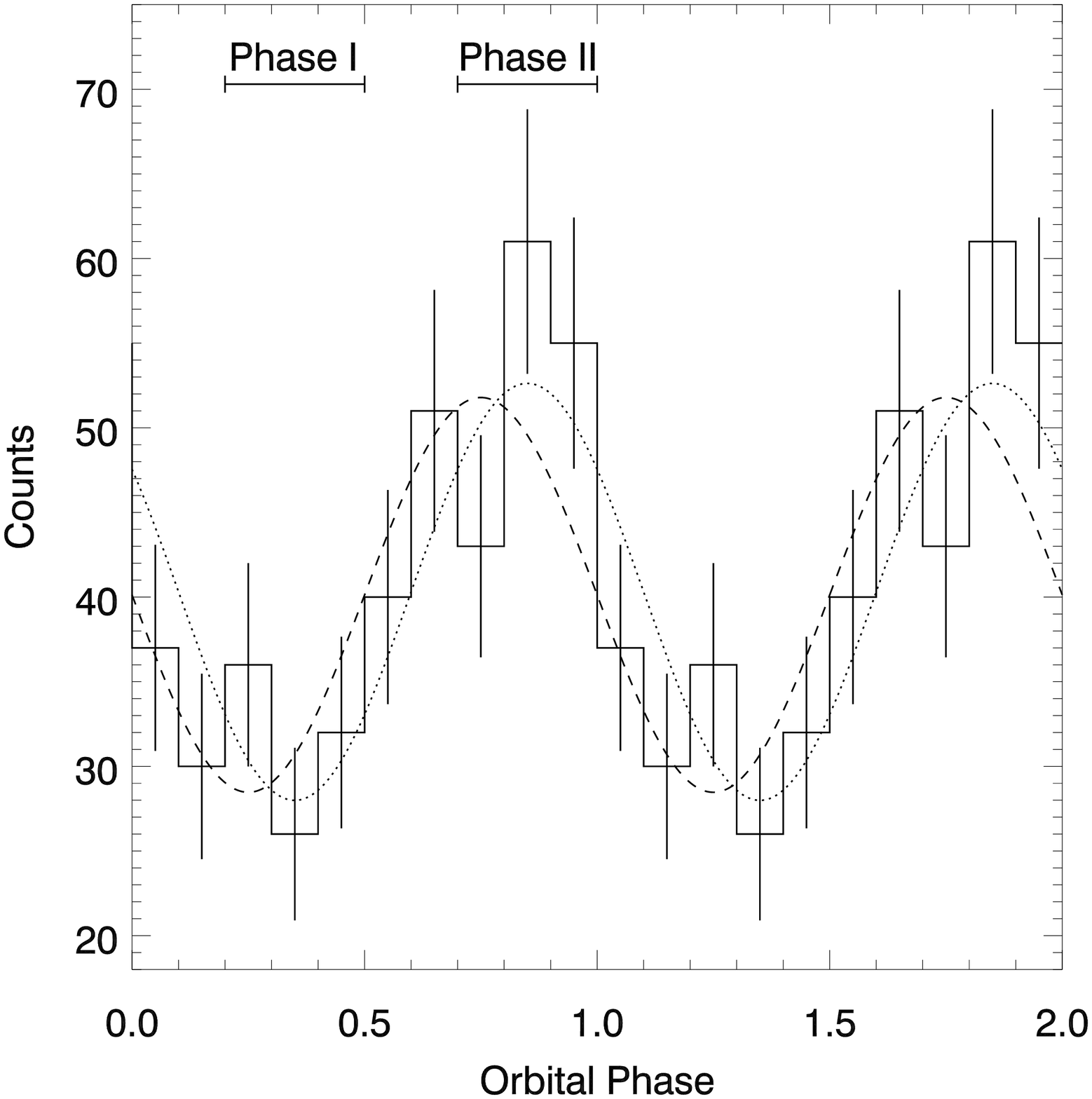}
\figcaption{0.2--300 GeV light curve folded at the orbital period using 
the offpulse data. The bottom (Phase I) and peak (Phase II) ranges of
the modulation are marked. Two simple sinusoid fits are displayed as
dashed and dotted curves; for the latter, 0.1 phase shift is forced
(see the text in Discussion).
\label{fig:orbital}}
\end{center}

\subsection{Maximum Likelihood Analysis}
\label{subsec:mla}

We selected LAT events in 0.1--300\,GeV energy range for the likelihood 
analysis, and included all sources within 20 deg in the \textit{Fermi} 
third source catalog \citep{ace+15} centered at the position of 
\psr\ to make the source model. 
The spectral function forms of the sources are provided in the catalog. 
The spectral parameters of the sources within 5 deg from \psr\ were 
set free, and all other parameters of the sources were fixed at their 
catalog values.  The \gr\ counterpart of \psr\ was modeled with an 
exponentially cutoff power law, characteristic for pulsars \citep{abd+13}, 
and a simple power law for comparison. In addition, we used the spectrum 
model gll\_iem\_v05\_rev1.fits and the spectrum file iso\_source\_v05.txt 
to consider the Galactic and extragalactic diffuse emission, respectively.
\begin{figure*}
\centering
\includegraphics[scale=0.28]{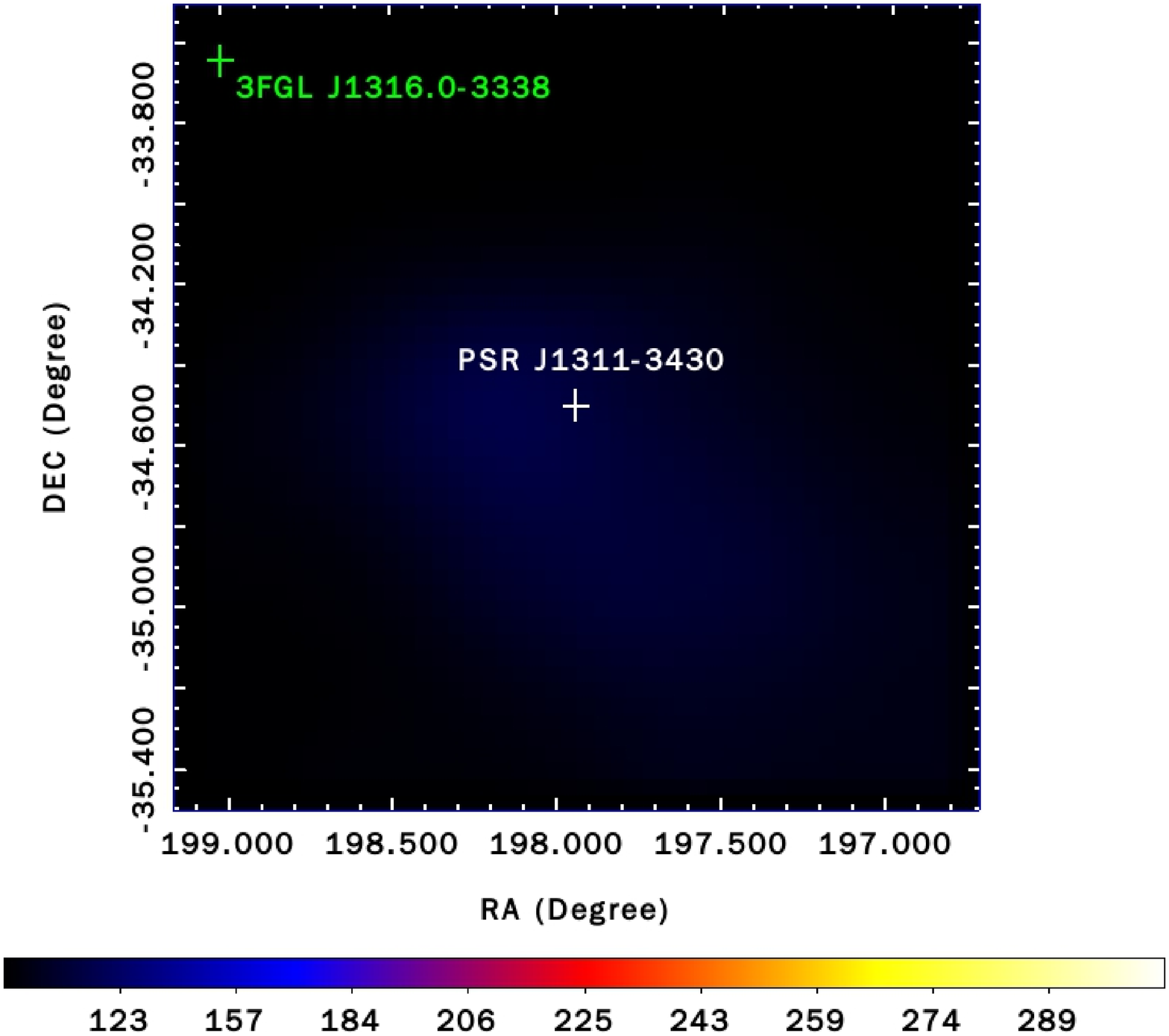}
\includegraphics[scale=0.28]{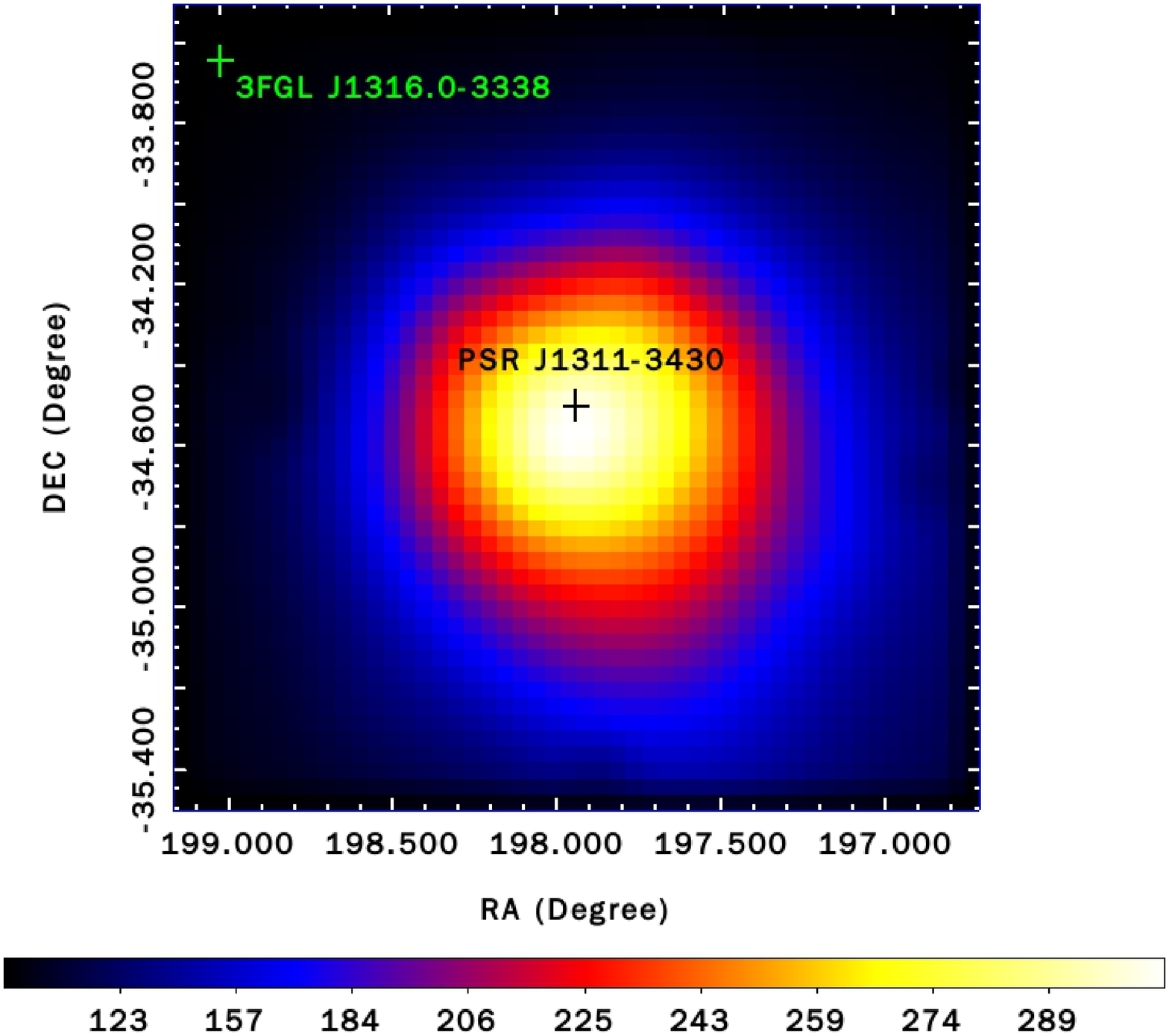}
\caption{0.2--300 GeV TS maps of a $\mathrm{2^{o}\times2^{o}}$ region 
centered at the position of \psr\ during Phase I ({\it left}) and 
Phase II ({\it right}). The image scale of the maps is 
0.04\arcdeg\ pixel$^{-1}$. The color bars indicate the TS value range. 
All sources in the source model except \psr\ were considered and removed. 
The white ({\it left}) and dark ({\it right}) crosses mark the position 
of \psr, and the green cross marks the position of the nearby catalog 
source 3FGL J1316.0$-$3338.} 
\label{fig:tsmap-orbital}
\end{figure*}

Using the LAT science tools software package v9r33p0, we performed
standard binned likelihood analysis to the LAT data. The \gr\ emission 
during the total pulse phase interval was detected with a Test Statistic (TS) 
value of 6279, while that during the onpulse and offpulse phase intervals 
were detected with TS values of 7723 and 499, respectively.
The TS value at a specific position is calculated from 
TS$= -2\log(L_{0}/L_{1})$, where $L_{0}$ 
and $L_{1}$ are the maximum likelihood values for a model without and with 
an additional source respectively, and approximately is the square of 
the detection significance for the additional source \citep{abd+10}.
We found during the total pulse phase, onpulse phase, and offpulse phase 
intervals, the emission is better modeled by an exponentially cutoff power 
law, with the low energy cutoff detected with $>$13$\sigma$, $>$14$\sigma$, 
and $>$5$\sigma$ significance (estimated from $\sqrt{-2\log(L_{pl}/L_{exp})}$, 
where $L_{exp}$ and $L_{pl}$ are the maximum likelihood values for 
the exponentially cutoff power-law model and power-law model, respectively;
\citealt{abd+13}). The resulting exponentially cutoff power-law fits 
are summarized in Table~\ref{tab:likelihood}.

\subsection{Orbital Variability}
\label{subsec:ov}

We folded the LAT events of the \psr\ region at its orbital period 
(\citealt{ple+12}) to study its possible orbital modulations. The
source position given in \citet{ple+12} was used for 
the barycentric corrections to photon arrival times, and photons 
within R$_{max}$ (R$_{max}$ ranges from 0\fdg1--1\fdg0 with a step of 0\fdg1) 
from the position were collected. 
Different energy ranges 
(0.1--300, 0.2--300, 0.3--300, 0.5-300, 1--300 GeV)
were tested in folding. No significant modulations were detected using 
the whole data (i.e., during the total pulse phase), as the largest H-test 
value was 6 (corresponding to $<$2$\sigma$ detection significance; 
\citealt{jrs89}). However, a significant orbital signal was best 
revealed using the offpulse data in the $>$0.2 GeV energy range 
within 0\fdg4 from \psr.  The folded light curve, which has
an H-test value of $\sim$22 (corresponding to $\sim$4$\sigma$
significance and $\sim$3$\sigma$ post-trial significance, for the
latter where 50 trials on the energy range and aperture radius are considered),
is shown in Figure~\ref{fig:orbital}. 
The phase zero is at the ascending node of the pulsar \citep{ple+12}. 
The folded light curve has a brightness peak around the superior conjunction 
(when the companion is behind the pulsar), the same as the modest X-ray one 
reported in \citet{r12}. The similarity helps strengthen the \gr\ modulation 
detection. Using the LAT tool {\tt gtexposure}, we checked the summed
exposures over the 10 orbital phase bins (e.g., \citealt{joh+15}), 
and they had only  $<$1\% differences, too small to cause any
artificial orbital modulations.

We performed likelihood analysis to the $>$0.2 GeV offpulse LAT events 
during the orbital phase ranges of 0.2--0.5 (named Phase I) and 0.7--1.0 
(named Phase II), which were approximately defined for the bottom and peak 
of the orbital modulation, respectively. We found that the emission 
during the both phase ranges is better modeled by an exponentially cutoff 
power law, with low energy cutoff detected with $>$3$\sigma$ and 
$>$2$\sigma$ significance. The exponentially cutoff power-law fits are 
summarized in Table~\ref{tab:likelihood}. The TS values during Phase I and II 
are $\sim$120 and $\sim$320 (see Figure~\ref{fig:tsmap-orbital}), 
respectively, which indicate that 
the source during the latter is more significantly detected than during 
the former, confirming the detection of orbital modulation from photon 
folding. 

Possible contamination from a nearby catalog source 3FGL J1316.0$-$3338,
which is identified as the counterpart to the flat spectrum radio quasar
(FSRQ) PKS 1313$-$333 \citep{ack+15}, was investigated. 
This source, being only $\sim$1\fdg2 away from \psr\ and relatively
bright (TS$\simeq$625 in the catalog), exhibited flaring events in the past\footnote{\footnotesize http://fermi.gsfc.nasa.gov/ssc/data/access/lat/4yr\_catalog/ap\_lcs.php}.
Using the offpulse data and performing likelihood analysis, we extracted 
its 30-day interval light curve, and found that
for five time bins (MJD 54863--54923, MJD 55433--55493, MJD 56363-56393), 
it had fluxes $>$2$\sigma$ above the value obtained from
the total offpulse data. We repeated the analysis by excluding the data of 
the time bins. We found that the folded light curve is nearly the same, 
still having an H-test value of $\simeq$22. 
\begin{figure*}
\centering
\includegraphics[scale=0.28]{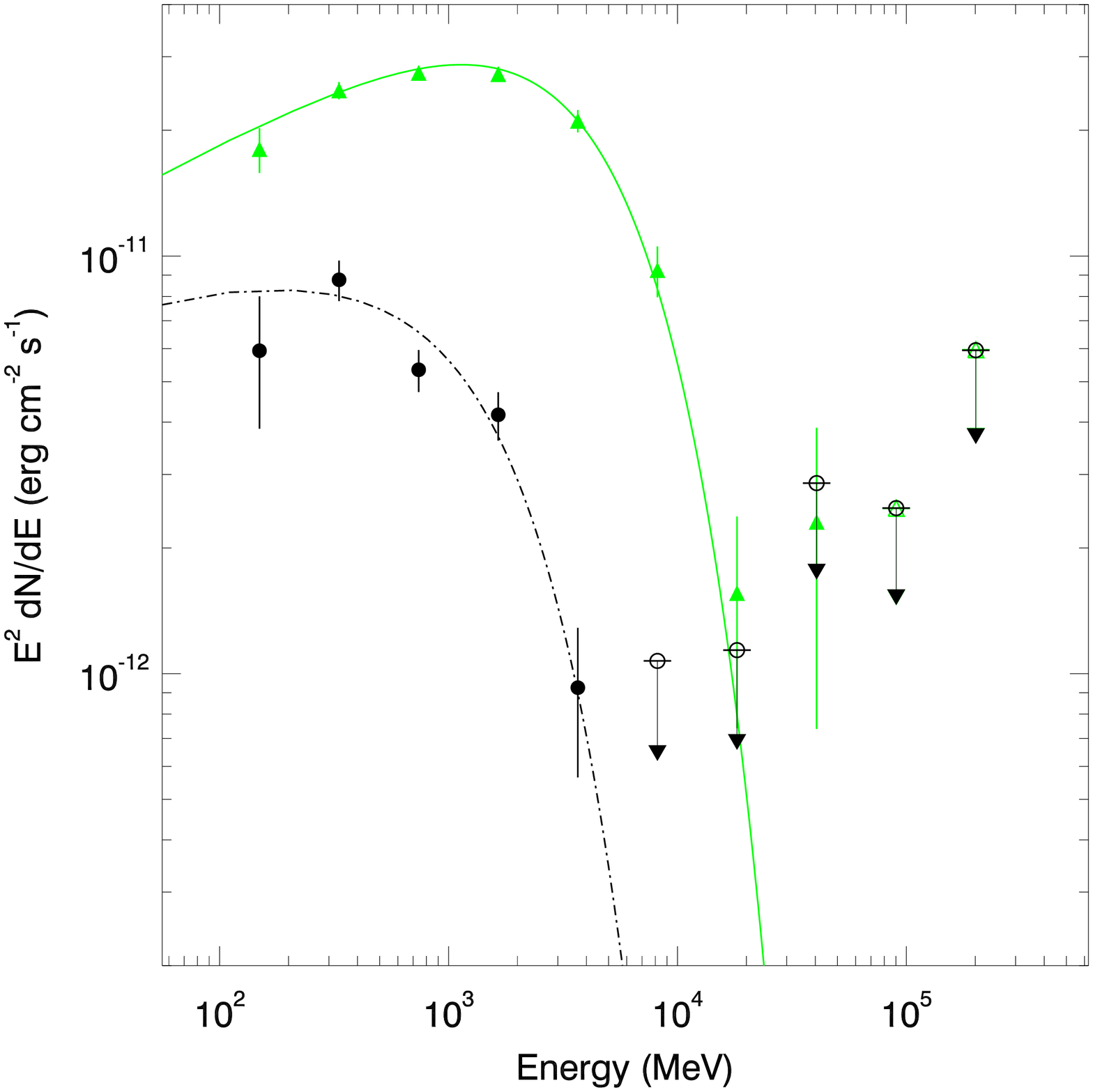}
\includegraphics[scale=0.28]{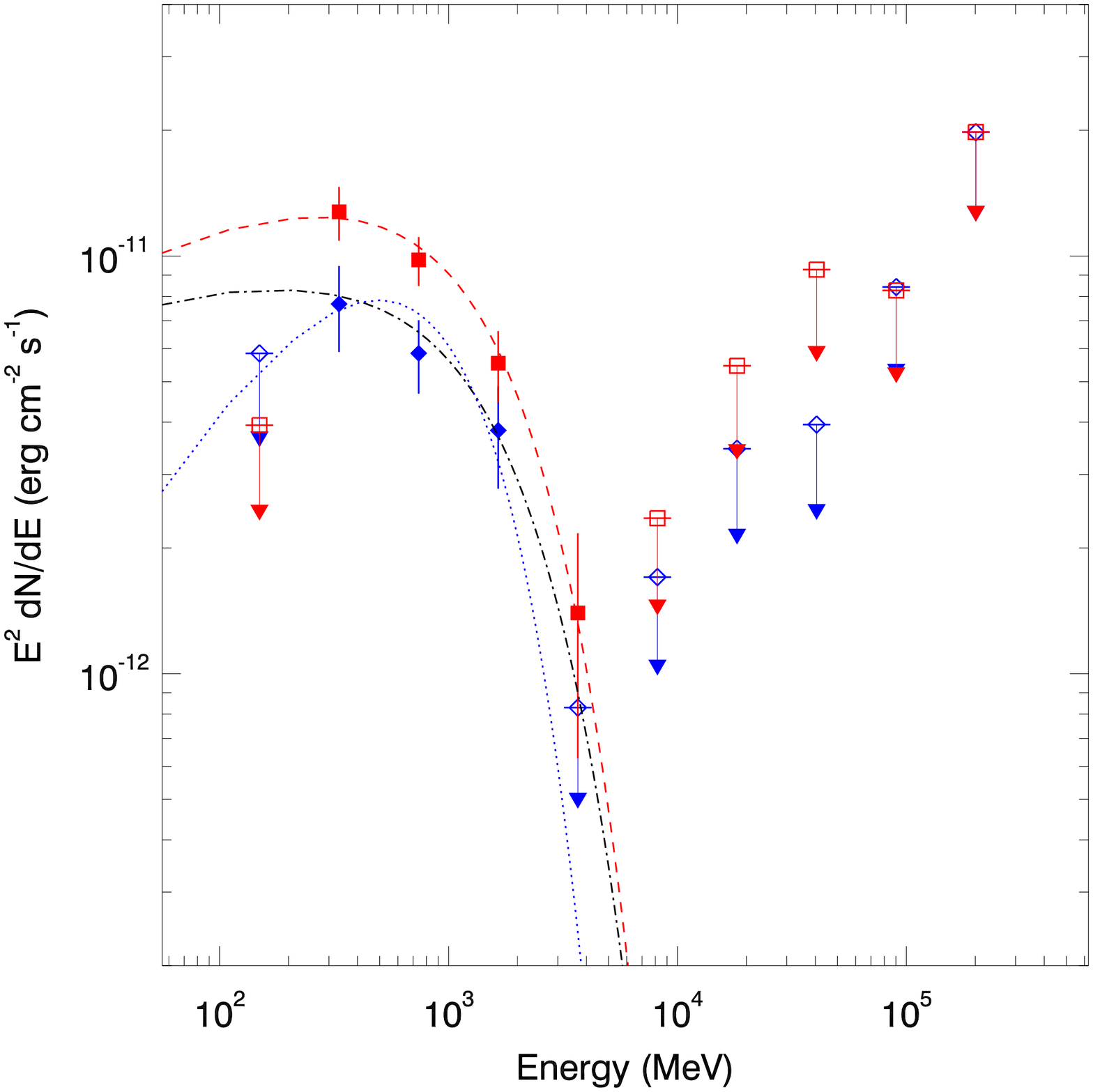}
\caption{{\it Left panel:} $\gamma$-ray spectra of \psr\ 
obtained with the onpulse (green triangles) and offpulse data (dark circles). 
{\it Right panel:} $\gamma$-ray spectra of \psr\ obtained during offpulse
Phase I (blue diamonds) and Phase II (red squares). 
The exponentially cutoff power-law fits
for the onpulse and offpulse data are shown as green solid and dark 
dot-dashed curves, respectively. The same model fits for the offpulse
Phase I and Phase II data are shown as blue dotted and red dashed curves, 
respectively.}
\label{fig:spectra}
\end{figure*}

\subsection{Spectral Analysis}
\label{subsec:sa}

We further investigated the orbital-dependent spectral variability 
during the offpulse phase interval. Spectra of \psr\ during the whole 
offpulse phase interval, Phase I, and Phase II were obtained, and the spectrum 
during the onpulse phase interval was also obtained for comparison. We
extracted the spectra by performing maximum likelihood analysis to the LAT data 
in 10 evenly divided energy bins in logarithm from 0.1--300 GeV, 
with the emission of the source being modeled with a power law in each energy 
bin. We only kept spectral points with TS$\geq$4, and derived 
the 95\% upper limits in other energy bins. The spectra extracted by this 
method are less model-dependent and provide a detailed description of 
the \gr\ emission for the source.

The obtained spectra are shown in Figure~\ref{fig:spectra}. The onpulse 
emission appears to have a 3-times higher cutoff energy $E_c$ 
than the offpulse one 
(see also Table~\ref{tab:likelihood}). In addition, comparing the two 
offpulse spectra, the source was brighter across 
the $>$0.2 GeV energy range during Phase II than during Phase I. 

We also repeated the analysis by excluding the data when 
the nearby source 3FGL~J1316.0$-$3338 had possible flares 
(see \S~\ref{subsec:ov}).
The obtained spectral parameters of \psr\ during the two 
orbital-phase ranges are consistent with the values obtained above (within 
1$\sigma$ uncertainties). We concluded that the flares do not have any 
significant effect on our spectral analysis. However, we note that 
in the lowest energy bin, the two upper limits of
the Phase I and II spectra (see the right panel
of Figure~\ref{fig:spectra}) are lower than the exponentially cutoff 
power-law fits. This problem might be due to possible contamination from 
3FGL~J1316.0$-$3338 in the low-energy range.

\section{Discussion}
\label{sec:dis}

From our analysis of the \textit{Fermi} data of \psr, we have detected 
its \gr\ orbital modulation during the offpulse phase interval of the pulsar,
where the magnetospheric emission from the pulsar was likely effectively 
removed. In both optical and X-ray observations, flares were detected
\citep{r12}, indicating the strong interaction between the pulsar wind
and the companion. Likely $\gamma$-rays are also produced due to the intrabinary
interaction. However, different from that is seen in PSR B1957$+$20,
which has an extra component above 2.7\,GeV at its inferior conjunction
(when the companion is in front of the pulsar), the light curve peak is near the
superior conjunction for \psr. The difference suggests that the intrabinary
\gr\ emission model, which explains the extra component as the result
from viewing an inverse Compton process as a head-on collision
(see also \cite{bed14}), proposed in \citet{wu+12} does not apply here.

For \psr, no significant spectral changes were found from the offpulse 
orbital-phase--resolved spectra (Figure~\ref{fig:spectra}). 
The source appeared brighter across the $>$0.2 GeV energy range during 
the peak range (Phase II) than during the bottom range (Phase I). 
Although the uncertainties from our fits with the exponentially cutoff power 
law are relatively large, the two spectra are generally
similar to each other. 
The similarity suggests an geometric origin for the orbital modulation,
such as that used to explain the X-ray orbital modulation of PSR J1023+0038
\citep{bog+11}. The binary likely has an inclination angle of $i\sim$60\arcdeg,
and the companion has a very small Roche lobe radius $R_L=0.068\ R_{\sun}$
(a canonical neutron star mass $M_n=1.35\ M_{\sun}$ is assumed). Therefore
like in PSR J1023+0038, the intrabinary emission region 
must be very close to the companion and the companion can thus block part
of the region, causing the observed orbital modulation. Here
we simply assume that the region is at the inner surface of the companion,
and use a function $m_h[1+\sin(2\pi\phi-\pi)\sin i]/2+m_c$ to describe 
the orbital
modulation, where $\phi$ is the orbital phase, $i=60\arcdeg$ is fixed,
and $m_h$ and $m_c$ are
the modulation amplitude (in units of counts) and constant counts, respectively.
Fitting the folded light curve, we found this function can describe 
the modulation (see Figure~\ref{fig:orbital}), where the minimum 
$\chi^2=9.7$ (for 8 degrees of freedom) and $m_h=27\pm7$, $m_c=27\pm3$. 
However, examining the light curve, the minimum and maximum may
have a 0.1 phase shift (i.e., they occur at phase 0.35 and 0.85,
respectively). If we force such a shift, the results are
$\chi^2=7.9$ (for 8 degrees of freedom) and $m_h=28\pm7$, $m_c=26\pm3$,
indeed slightly better. This shift may suggest that the emission is not 
isotropic. The constant part $m_c$ may represent the intrabinary emission 
unblocked by the companion over the whole orbital phase, while emission 
from the pulsar during its 
offpulse phase interval could also contribute a small fraction.

It is not clear why \psr\ has an orbital modulation different from that
of PSR B1957+20. We note that if the size of the interaction region
is proportional to that of a companion, similar fractions of an isotropic
pulsar wind (0.0021 vs. 0.0024) would be intercepted 
for \psr\ and
PSR B1957+20 respectively (estimated from $(R_2/D_b)^2/4$; $R_2$ 
is the radius of the companion and $D_b$ is the separation distance 
of the binary). The notable differences are that the spin-down luminosity of
\psr\ is approximately 1/3 of PSR B1957+20 and that the companion in
\psr\  nearly fills its Roche lobe \citep{rom+12}. We suspect that 
because of the latter, an outflow from the companion may still exist, the
same as that in PSR J1023+0038 (e.g., \citealt{bog+11}).
For PSR B1957+20, its companion fills its Roche lobe 
$\sim 85\%$ \citep{rey+07}, and the mass loss from the companion
is presumably driven by the pulsar wind (\citealt{at93} and references therein).
The much more messy environment in \psr, as suggested by the radio observations
\citep{ray+13}, could be evidence for this possibility. Detailed modeling
for physical processes in \psr\ would help verify it.

Our discovery of the orbital \gr\ modulation in \psr\ and the analysis
results have provided clear evidence for \gr\ production due to intrabinary
interaction between a pulsar and its companion in a black widow system, 
and thus have confirmed the general physical picture that has long been proposed
theoretically. Since X-ray observations have revealed the general existence
of intrabinary shock emission in eclipsing MSP binaries, similar 
work can be carried out to search and study related \gr\ emission from
these recently identified systems.

\acknowledgements
We thank the anonymous referee for helpful suggestions.
This research made use of the High Performance Computing Resource in the Core
Facility for Advanced Research Computing at Shanghai Astronomical Observatory.
This research was supported by the Shanghai Natural Science 
Foundation for Youth (13ZR1464400), the National Natural Science Foundation
of China for Youth (11403075), the National Natural Science Foundation
of China (11373055), and the Strategic Priority Research Program
``The Emergence of Cosmological Structures" of the Chinese Academy
of Sciences (Grant No. XDB09000000). Z.W. is a Research Fellow of the
One-Hundred-Talents project of Chinese Academy of Sciences.

\bibliographystyle{apj}

\begin{thebibliography}{30}
\expandafter\ifx\csname natexlab\endcsname\relax\def\natexlab#1{#1}\fi

\bibitem[{{Abdo} {et~al.}(2010){Abdo}, {Ackermann}, {Ajello}, {Allafort},
  {Antolini}, {Atwood}, {Axelsson}, {Baldini}, {Ballet}, {Barbiellini}, \&
  et~al.}]{abd+10}
{Abdo}, A.~A., {et~al.} 2010, \apjs, 188, 405

\bibitem[{{Abdo} {et~al.}(2013){Abdo}, {Ajello}, {Allafort}, {Baldini},
  {Ballet}, {Barbiellini}, {Baring}, {Bastieri}, {Belfiore}, {Bellazzini}, \&
  et~al.}]{abd+13}
---. 2013, \apjs, 208, 17

\bibitem[{{Ackerman} {et~al.}(2015)}]{ack+15}
Ackerman, M., Ajello, M., Atwood, W., et al. 2015, ArXiv:1501.06054

\bibitem[{{Alpar} {et~al.}(1982){Alpar}, {Cheng}, {Ruderman}, \&
  {Shaham}}]{alp+82}
{Alpar}, M.~A., {Cheng}, A.~F., {Ruderman}, M.~A., \& {Shaham}, J. 1982, \nat,
  300, 728

\bibitem[{{Archibald} {et~al.}(2010){Archibald}, {Kaspi}, {Bogdanov},
  {Hessels}, {Stairs}, {Ransom}, \& {McLaughlin}}]{arc+10}
{Archibald}, A.~M., {Kaspi}, V.~M., {Bogdanov}, S., {Hessels}, J.~W.~T.,
  {Stairs}, I.~H., {Ransom}, S.~M., \& {McLaughlin}, M.~A. 2010, \apj, 722, 88

\bibitem[{{Arons} \& {Tavani}(1993)}]{at93}
{Arons}, J., \& {Tavani}, M. 1993, \apj, 403, 249

\bibitem[{{Bednarek}(2014)}]{bed14}
{Bednarek}, W. 2014, \aap, 561, A116

\bibitem[{{Bogdanov} {et~al.}(2011){Bogdanov}, {Archibald}, {Hessels}, {Kaspi},
  {Lorimer}, {McLaughlin}, {Ransom}, \& {Stairs}}]{bog+11}
{Bogdanov}, S., {Archibald}, A.~M., {Hessels}, J.~W.~T., {Kaspi}, V.~M.,
  {Lorimer}, D., {McLaughlin}, M.~A., {Ransom}, S.~M., \& {Stairs}, I.~H. 2011,
  \apj, 742, 97

\bibitem[{{Bogdanov} {et~al.}(2014){Bogdanov}, {Patruno}, {Archibald}, {Bassa},
  {Hessels}, {Janssen}, \& {Stappers}}]{bog+14}
{Bogdanov}, S., {Patruno}, A., {Archibald}, A.~M., {Bassa}, C., {Hessels},
  J.~W.~T., {Janssen}, G.~H., \& {Stappers}, B.~W. 2014, \apj, 789, 40

\bibitem[{{de Jager} {et~al.}(1989){de Jager}, {Raubenheimer}, \&
  {Swanepoel}}]{jrs89}
{de Jager}, O.~C., {Raubenheimer}, B.~C., \& {Swanepoel}, J.~W.~H. 1989, \aap,
  221, 180

\bibitem[{{Edwards} {et~al.}(2006){Edwards}, {Hobbs}, \& {Manchester}}]{ehm06}
{Edwards}, R.~T., {Hobbs}, G.~B., \& {Manchester}, R.~N. 2006, \mnras, 372,
  1549

\bibitem[{{Fruchter} {et~al.}(1988){Fruchter}, {Stinebring}, \&
  {Taylor}}]{fst88}
{Fruchter}, A.~S., {Stinebring}, D.~R., \& {Taylor}, J.~H. 1988, \nat, 333, 237

\bibitem[{{Gentile} {et~al.}(2014){Gentile}, {Roberts}, {McLaughlin}, {Camilo},
  {Hessels}, {Kerr}, {Ransom}, {Ray}, \& {Stairs}}]{gen+14}
{Gentile}, P.~A., {et~al.} 2014, \apj, 783, 69

\bibitem[{{Hobbs} {et~al.}(2006){Hobbs}, {Edwards}, \& {Manchester}}]{hem06}
{Hobbs}, G.~B., {Edwards}, R.~T., \& {Manchester}, R.~N. 2006, \mnras, 369, 655

\bibitem[{{Huang} {et~al.}(2012){Huang}, {Kong}, {Takata}, {Hui}, {Lin}, \&
  {Cheng}}]{hua+12}
{Huang}, R.~H.~H., {Kong}, A.~K.~H., {Takata}, J., {Hui}, C.~Y., {Lin},
  L.~C.~C., \& {Cheng}, K.~S. 2012, \apj, 760, 92

\bibitem[{{Kataoka} {et~al.}(2012){Kataoka}, {Yatsu}, {Kawai}, {Urata},
  {Cheung}, {Takahashi}, {Maeda}, {Totani}, {Makiya}, {Hanayama}, {Miyaji}, \&
  {Tsai}}]{kat+12}
{Kataoka}, J., {et~al.} 2012, \apj, 757, 176

\bibitem[{{Manchester} {et~al.}(2005){Manchester}, {Hobbs}, {Teoh}, \&
  {Hobbs}}]{man+05}
{Manchester}, R.~N., {Hobbs}, G.~B., {Teoh}, A., \& {Hobbs}, M. 2005, \aj, 129,
  1993

\bibitem[{{Nolan} {et~al.}(2012){Nolan}, {Abdo}, {Ackermann}, {Ajello},
  {Allafort}, {Antolini}, {Atwood}, {Axelsson}, {Baldini}, {Ballet}, \&
  et~al.}]{nol+12}
{Nolan}, P.~L., {et~al.} 2012, \apjs, 199, 31

\bibitem[{{Pletsch} {et~al.}(2012){Pletsch}, {Guillemot}, {Fehrmann}, {Allen},
  {Kramer}, {Aulbert}, {Ackermann}, {Ajello}, {de Angelis}, {Atwood},
  {Baldini}, {Ballet}, {Barbiellini}, {Bastieri}, {Bechtol}, {Bellazzini},
  {Borgland}, {Bottacini}, {Brandt}, {Bregeon}, {Brigida}, {Bruel}, {Buehler},
  {Buson}, {Caliandro}, {Cameron}, {Caraveo}, {Casandjian}, {Cecchi}, {{\c
  C}elik}, {Charles}, {Chaves}, {Cheung}, {Chiang}, {Ciprini}, {Claus},
  {Cohen-Tanugi}, {Conrad}, {Cutini}, {D'Ammando}, {Dermer}, {Digel}, {Drell},
  {Drlica-Wagner}, {Dubois}, {Dumora}, {Favuzzi}, {Ferrara}, {Franckowiak},
  {Fukazawa}, {Fusco}, {Gargano}, {Gehrels}, {Germani}, {Giglietto},
  {Giordano}, {Giroletti}, {Godfrey}, {Grenier}, {Grondin}, {Grove}, {Guiriec},
  {Hadasch}, {Hanabata}, {Harding}, {den Hartog}, {Hayashida}, {Hays}, {Hill},
  {Hou}, {Hughes}, {J{\'o}hannesson}, {Jackson}, {Jogler}, {Johnson},
  {Johnson}, {Kataoka}, {Kerr}, {Kn{\"o}dlseder}, {Kuss}, {Lande}, {Larsson},
  {Latronico}, {Lemoine-Goumard}, {Longo}, {Loparco}, {Lovellette}, {Lubrano},
  {Massaro}, {Mayer}, {Mazziotta}, {McEnery}, {Mehault}, {Michelson},
  {Mitthumsiri}, {Mizuno}, {Monzani}, {Morselli}, {Moskalenko}, {Murgia},
  {Nakamori}, {Nemmen}, {Nuss}, {Ohno}, {Ohsugi}, {Omodei}, {Orienti},
  {Orlando}, {de Palma}, {Paneque}, {Perkins}, {Piron}, {Pivato}, {Porter},
  {Rain{\`o}}, {Rando}, {Ray}, {Razzano}, {Reimer}, {Reimer}, {Reposeur},
  {Ritz}, {Romani}, {Romoli}, {Sanchez}, {Parkinson}, {Schulz}, {Sgr{\`o}}, {do
  Couto e Silva}, {Siskind}, {Smith}, {Spandre}, {Spinelli}, {Suson},
  {Takahashi}, {Tanaka}, {Thayer}, {Thayer}, {Thompson}, {Tibaldo},
  {Tinivella}, {Troja}, {Usher}, {Vandenbroucke}, {Vasileiou}, {Vianello},
  {Vitale}, {Waite}, {Winer}, {Wood}, {Wood}, {Yang}, \& {Zimmer}}]{ple+12}
{Pletsch}, H.~J., {et~al.} 2012, Science, 338, 1314

\bibitem[{{Radhakrishnan} \& {Srinivasan}(1982)}]{rs82}
{Radhakrishnan}, V., \& {Srinivasan}, G. 1982, Current Science, 51, 1096

\bibitem[{{Ray} {et~al.}(2013){Ray}, {Ransom}, {Cheung}, {Giroletti},
  {Cognard}, {Camilo}, {Bhattacharyya}, {Roy}, {Romani}, {Ferrara},
  {Guillemot}, {Johnston}, {Keith}, {Kerr}, {Kramer}, {Pletsch}, {Saz
  Parkinson}, \& {Wood}}]{ray+13}
{Ray}, P.~S., {et~al.} 2013, \apjl, 763, L13

\bibitem[{{Reynolds} {et~al.}(2007){Reynolds}, {Callanan}, {Fruchter},
  {Torres}, {Beer}, \& {Gibbons}}]{rey+07}
{Reynolds}, M.~T., {Callanan}, P.~J., {Fruchter}, A.~S., {Torres}, M.~A.~P.,
  {Beer}, M.~E., \& {Gibbons}, R.~A. 2007, \mnras, 379, 1117

\bibitem[{{Roberts}(2013)}]{r13}
{Roberts}, M.~S.~E. 2013, in IAU Symposium, Vol. 291, IAU Symposium, ed.
  J.~{van Leeuwen}, 127--132

\bibitem[{{Roberts} {et~al.}(2014){Roberts}, {Mclaughlin}, {Gentile}, {Aliu},
  {Hessels}, {Ransom}, \& {Ray}}]{rob+14}
{Roberts}, M.~S.~E., {Mclaughlin}, M.~A., {Gentile}, P., {Aliu}, E., {Hessels},
  J.~W.~T., {Ransom}, S.~M., \& {Ray}, P.~S. 2014, Astronomische Nachrichten,
  335, 313

\bibitem[{{Romani}(2012)}]{r12}
{Romani}, R.~W. 2012, \apjl, 754, L25

\bibitem[{{Romani} {et~al.}(2012){Romani}, {Filippenko}, {Silverman}, {Cenko},
  {Greiner}, {Rau}, {Elliott}, \& {Pletsch}}]{rom+12}
{Romani}, R.~W., {Filippenko}, A.~V., {Silverman}, J.~M., {Cenko}, S.~B.,
  {Greiner}, J., {Rau}, A., {Elliott}, J., \& {Pletsch}, H.~J. 2012, \apjl,
  760, L36

\bibitem[{{Johnson} {et~al.}(2015)}]{joh+15}
{Johnson}, T.~J., {Ray}, P.~S., {Roy}, J., {Cheung}, C.~C.,
    {Harding}, A.~K., {Pletsch}, H.~J., {Fort}, S., {Camilo}, F.,
    {Deneva}, J., {Bhattacharyya}, B., {Stappers}, B.~W. 2015, ArXiv:1502.06862

\bibitem[{{Taylor}(1992)}]{t92}
{Taylor}, J.~H. 1992, Royal Society of London Philosophical Transactions Series
  A, 341, 117

\bibitem[{{The Fermi-LAT Collaboration}(2015)}]{ace+15}
{The Fermi-LAT Collaboration}. 2015, ArXiv e-prints

\bibitem[{{Wu} {et~al.}(2012){Wu}, {Takata}, {Cheng}, {Huang}, {Hui}, {Kong},
  {Tam}, \& {Wu}}]{wu+12}
{Wu}, E.~M.~H., {Takata}, J., {Cheng}, K.~S., {Huang}, R.~H.~H., {Hui}, C.~Y.,
  {Kong}, A.~K.~H., {Tam}, P.~H.~T., \& {Wu}, J.~H.~K. 2012, \apj, 761, 181

\bibitem[{{Xing} \& {Wang}(2014)}]{xw14}
{Xing}, Y., \& {Wang}, Z. 2014, ArXiv e-prints

\bibitem[{{Xing} {et~al.}(2014){Xing}, {Wang}, \& {Ng}}]{xwn14}
{Xing}, Y., {Wang}, Z., \& {Ng}, C.-Y. 2014, \apj, 795, 88

\end{thebibliography}

\begin{deluxetable}{lcccc}
\tablecaption{Exponentially cutoff power-law fits for \psr.}
\tablewidth{0pt}
\startdata
\hline
\hline
Data set & $>$0.1 GeV Flux & $\Gamma$ & E$_{c}$ & TS \\
 & (10$^{-8}$ photon cm$^{-2}$ s$^{-1}$) &  & (GeV) &   \\
\hline
Total data & 8.9 $\pm$ 0.4 & 1.80 $\pm$ 0.04 & 4.0 $\pm$ 0.5 & 6279 \\
\hline
Onpulse data & 13.9 $\pm$ 0.6 & 1.71 $\pm$ 0.04 & 3.9 $\pm$ 0.4 & 7723 \\
\hline
Offpulse data & 4.6 $\pm$ 0.6 & 1.9 $\pm$ 0.2 & 1.3 $\pm$ 0.4 & 499 \\
\hline
Offpulse Phase I & 3 $\pm$ 1 & 1.2 $\pm$ 0.7 & 0.6 $\pm$ 0.3 & 120 \\
\hline
Offpulse Phase II & 7 $\pm$ 2 & 1.8 $\pm$ 0.5 & 1.2 $\pm$ 0.9 & 320 \\
\enddata
\tablecomments{Column 3 and 4 list the photon index and cutoff energy of the exponentially cutoff power-law model.}
\label{tab:likelihood}
\end{deluxetable}

\end{document}